\documentclass[aps,prx,twocolumn,superscriptaddress,floatfix]{revtex4-1}
\pdfoutput=1

\usepackage{amsmath,epsfig,color,amssymb}
\usepackage{dcolumn}
\usepackage{bm}
\usepackage{graphicx}
\usepackage{amsmath, amsfonts, amssymb,mathrsfs}
\usepackage{pstricks}
\usepackage{amsxtra}
\usepackage{amsthm}
\usepackage{natbib}
\usepackage{color}
\usepackage{float}
\usepackage{epstopdf}
\usepackage{makecell}
\usepackage{layouts}
\usepackage{longtable}
\usepackage[english]{babel}

\makeatletter
\let\save@mathaccent\mathaccent
\newcommand*\if@single[3]{%
  \setbox0\hbox{${\mathaccent"0362{#1}}^H$}%
  \setbox2\hbox{${\mathaccent"0362{\kern0pt#1}}^H$}%
  \ifdim\ht0=\ht2 #3\else #2\fi
  }
\newcommand*\rel@kern[1]{\kern#1\dimexpr\macc@kerna}
\newcommand*\widebar[1]{\@ifnextchar^{{\wide@bar{#1}{0}}}{\wide@bar{#1}{1}}}
\newcommand*\wide@bar[2]{\if@single{#1}{\wide@bar@{#1}{#2}{1}}{\wide@bar@{#1}{#2}{2}}}
\newcommand*\wide@bar@[3]{%
  \begingroup
  \def\mathaccent##1##2{%
    \let\mathaccent\save@mathaccent
    \if#32 \let\macc@nucleus\first@char \fi
    \setbox\z@\hbox{$\macc@style{\macc@nucleus}_{}$}%
    \setbox\tw@\hbox{$\macc@style{\macc@nucleus}{}_{}$}%
    \dimen@\wd\tw@
    \advance\dimen@-\wd\z@
    \divide\dimen@ 3
    \@tempdima\wd\tw@
    \advance\@tempdima-\scriptspace
    \divide\@tempdima 10
    \advance\dimen@-\@tempdima
    \ifdim\dimen@>\z@ \dimen@0pt\fi
    \rel@kern{0.6}\kern-\dimen@
    \if#31
      \overline{\rel@kern{-0.6}\kern\dimen@\macc@nucleus\rel@kern{0.4}\kern\dimen@}%
      \advance\dimen@0.4\dimexpr\macc@kerna
      \let\final@kern#2%
      \ifdim\dimen@<\z@ \let\final@kern1\fi
      \if\final@kern1 \kern-\dimen@\fi
    \else
      \overline{\rel@kern{-0.6}\kern\dimen@#1}%
    \fi
  }%
  \macc@depth\@ne
  \let\math@bgroup\@empty \let\math@egroup\macc@set@skewchar
  \mathsurround\z@ \frozen@everymath{\mathgroup\macc@group\relax}%
  \macc@set@skewchar\relax
  \let\mathaccentV\macc@nested@a
  \if#31
    \macc@nested@a\relax111{#1}%
  \else
    \def\gobble@till@marker##1\endmarker{}%
    \futurelet\first@char\gobble@till@marker#1\endmarker
    \ifcat\noexpand\first@char A\else
      \def\first@char{}%
    \fi
    \macc@nested@a\relax111{\first@char}%
  \fi
  \endgroup
}
\makeatother
\makeatletter
\@ifundefined{textcolor}{}
{%
 \definecolor{BLACK}{gray}{0}
 \definecolor{WHITE}{gray}{1}
 \definecolor{RED}{rgb}{1,0,0}
 \definecolor{GREEN}{rgb}{0,1,0}
 \definecolor{BLUE}{rgb}{0,0,1}
 \definecolor{CYAN}{cmyk}{1,0,0,0}
 \definecolor{MAGENTA}{cmyk}{0,1,0,0}
 \definecolor{YELLOW}{cmyk}{0,0,1,0}
}

\def\be{\begin{equation}}
\def\ee{\end{equation}}
\def\bea{\begin{eqnarray}}
\def\eea{\end{eqnarray}}

\def\be{\begin{equation}}      
\def\ee{\end{equation}}
\def\beu{\begin{equation*}}   
\def\eeu{\end{equation*}}

\providecommand{\abs}[1]{\left\lvert#1\right\rvert}   
\providecommand{\ket}[1]{\left|#1\right\rangle}

\providecommand{\bra}[1]{\left\langle#1\right|}

\definecolor{new}{rgb}{.08,.05,.8}

\newcommand{\delete}[1]{{}} 

\graphicspath{}

\begin{document}
\title{Serialized Quantum Error Correction Protocol for High-Bandwidth Quantum Repeaters}
\author{A.~N.~Glaudell}
\affiliation{Joint Quantum Institute, University of Maryland, and the National Institute of Standards and Technology, College Park, MD 20742, USA}
\affiliation{Joint Center for Quantum Information and Computer Science, University of Maryland, College Park, MD 20742, USA}
\author{E.~Waks}
\affiliation{Joint Quantum Institute, University of Maryland, and the National Institute of Standards and Technology, College Park, MD 20742, USA}
\affiliation{Department of Electrical and Computer Engineering, Institute for Research in Electronics and
Applied Physics, University of Maryland, College Park, Maryland
20742, USA}
\author{J.~M.~Taylor}
\affiliation{Joint Quantum Institute, University of Maryland, and the National Institute of Standards and Technology, College Park, MD 20742, USA}
\affiliation{Joint Center for Quantum Information and Computer Science, University of Maryland, College Park, MD 20742, USA}

\date{\today}
\begin{abstract}
Advances in single photon creation, transmission, and detection suggest that sending quantum information over optical fibers may have losses low enough to be correctable using a quantum error correcting code. Such error-corrected communication is equivalent to a novel quantum repeater scheme, but crucial questions regarding implementation and system requirements remain open. Here we show that long range entangled bit generation with rates approaching $10^8$ ebits/s may be possible using a completely serialized protocol, in which photons are generated, entangled, and error corrected via sequential, one-way interactions with a minimal number of matter qubits. Provided loss and error rates of the required elements are below the threshold for quantum error correction, this scheme demonstrates improved performance over transmission of single photons. We find improvement in ebit rates at large distances using this serial protocol and various quantum error correcting codes.
\end{abstract}
\maketitle

\section{Introduction}
Quantum communication, from quantum key distribution \cite{Bennett1993,Ekert1991,Lo1999,Shor2000,Hwang2003} to distributed quantum computing \cite{Cirac1999,Serafini2006}, examines how entanglement between distant particles enables new information applications. Photons provide a natural means for sharing quantum information across long distances. However, optical fibers attenuate photon transmission, reducing entanglement generation rates exponentially with distance. Quantum repeaters \cite{Briegel1998a,Bennett1996a,Deutsch1996} use nearby entangled pairs of qubits (Bell pairs) to create longer-range entangled pairs via entanglement swapping and nested purification \cite{Dur1999}. Further improvements are possible \cite{Jiang2009} by using Bell pairs to purify qubits encoded in a quantum error correcting code (QECC) \cite{Shor1995,Steane1996}. However, these schemes suffer from low bit rates limited by the speed of light \cite{Jiang2007a}. When losses and other errors are below the threshold for quantum error correction \cite{Steane2003,Knill2005a}, a different approach emerges \cite{Fowler2010,Munro2010,Munro2012,Muralidharan2015}: photons can be sent as parts of encoded states of a QECC, and photon loss is recovered via quantum error correction. Crucially, the probability that a photon is lost from the creation to detection stage must be lower than the threshold of the associated QECC.

Motivated by recent advances in high efficiency detectors \cite{Miller2003,Lita2008,Marsili2013} and high efficiency cavity-fiber coupling \cite{Spillane2003,Barclay2005,Peter2005,Dousse2008,Thompson2013,Yu2014}, here we consider a minimalistic approach to produce logical states and correct errors in photonic systems with the aim of building a high bit rate quantum repeater. We assume continuing progress in recent demonstrations of single photon nonlinear gates \cite{Peyronel2012,Bose2012,Gao2012,Gao2013,Kim2013,Beterov2013,Pritchard2014,Loew2014,Tiecke2014,Sun2015} between light and matter qubits, and consider the limits for quantum error corrected quantum repeaters when we simultaneously minimize the number of matter qubits required, make few assumptions about long-term matter qubit coherence, and require that each photon interacts sequentially, and singly, with each matter qubit. We call this design a serial quantum repeater, and show in this paper that with current or near-future performance, one-way quantum communication at 1000 km ranges may approach the giga-entangled bit/second range using a stream of entangled photons in a narrow bandwidth over single fibers.

The signal for our serial repeater is comprised of several photons encoded in a codeword of a QECC; we use either polarization qubits or time-bin based qudits for the photons to distinguish loss errors from other errors. Our goal is to implement QECC-based teleportation of this signal to a new set of photons with sequential, single-shot interactions using stationary (matter) qudits. We will further restrict ourselves to using a cavity QED-based controlled phase gate \cite{Kim2013,Sun2015,Pritchard2014,Tiecke2014}, arbitrary local operations and measurements on single matter qudits, and Hadamard-type photon gates. These choices are designed to be consistent with recent progress in single-photon phase gates and on-chip photonics.

\begin{figure*}
\includegraphics{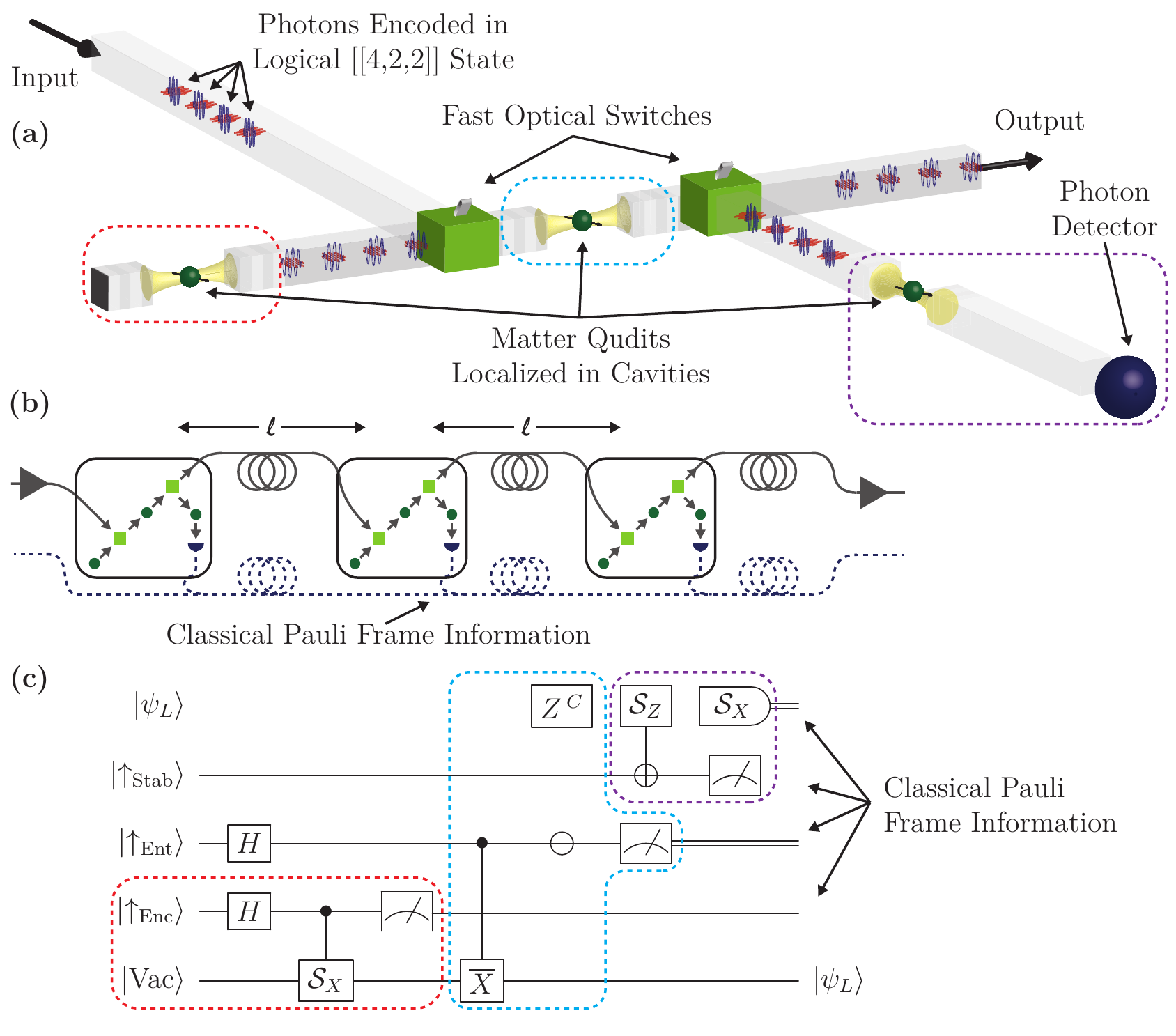}
\caption{\label{f:scheme}
(a) Example of a serialized QECC-based quantum repeater node for $[[4,2,2]]$ encoded photonic qubits in a polarization basis (time binning may also be used). Single photons are generated and entangled using a single quantum dot in an optical cavity (red box), with each sent after generation to be entangled with a quantum dot (blue box) acting as a short-term quantum memory for teleportation. The photons then continue to the output, heading to the next repeater node. Then the two optical switches (built, e.g., using on-chip Mach-Zender interferometers with fast phase modulators) are flipped. The incoming encoded photons entangle with the short-term memory (blue box), which is followed by stabilizer detection via interaction with a quantum dot (purple box) and destructive measurement using a multi-plexed, high efficiency beam-splitter network and photodetectors (abstractly shown with the blue half-ball).
(b) Multiple repeater nodes connected with optical fiber of length $\ell$. Additional Pauli frame information and error correction information are sent along the classical channel and aggregated for final correction at the end node of the repeater system.
(c) Quantum circuit for QECC-based teleportation of incoming photons carrying logical state $\ket{\psi_L}$ via serial interaction with three sets of matter qudits and newly created photons from $\ket{\textrm{Vac}}$. The outgoing state is the error corrected incoming logical state when errors are sufficiently low. Individual qudits are not shown; lines represent sets of qudits as described in the main text.
}
\end{figure*}

The procedure for implementing serial teleportation based error correction involves replacing each portion of the protocol for state teleportation \cite{Bennett1993,Bennett1996,Bennett1996a} with a new component commensurate with a QECC that keeps serialization intact. Incoming and outgoing states of photons are replaced by logical codewords of the QECC, which amounts to selecting how to represent the $k$ logical qudits with the $D^k$ codewords of the QECC.

Just before the arrival of the signal photons at a repeater node, the node will first generate a fresh set of photons encoded in a codeword of the QECC using a small set of matter qudits as single photon sources and for controlled phase gates. These outgoing photons are then entangled with a matter qudit `memory'. The outgoing photons are released toward the next node, while a switch inside the repeater enables incoming signal photons to now interact with the same matter qudit. After the interaction, measurement of the matter qudit teleports the signal state onto the outgoing photon state. However, the signal photons must also have their QECC stabilizers measured and losses detected. For CSS codes it suffices to first measure the $Z$-based stabilizers using an additional matter qudit per stabilizer, then apply a Hadamard to the photons and measure them with high-efficiency photon detectors. For an $[[n,k,d]]_D$ CSS code, $<(n-k)$ qudits are necessary for the encoding stage, from 1 to $k$ are necessary for the entangling stage, and $<(n-k)$ are necessary for the $Z$-stabilizer readout. The CSS codes considered in this work require only $(n-k)/2$ qudits for both encoding and stabilizer readout. In particular, for the minimal $[[3,1,2]]_3$ and $[[4,2,2]]$ codes, only three matter qudits are necessary for a completely serial approach.

Although we envision many physical implementations for the repeater architecture, here we consider the specific case of a single atomic-like  lambda system in a cavity, as demonstrated in a number of experiments \cite{Kim2013,Tiecke2014,Sun2015,Reiserer2013,Reiserer2014}.  In this case, the switching contrast -- bounding the CPHASE fidelity -- is given by $F = C^2/(1+C)^2$ where $C=2g^2/\gamma\kappa$ is the atomic cooperativity, $g$ is the cavity-quantum dot coupling strength, $\gamma$ is the atomic dipole decay rate, and $\kappa$ is the cavity energy decay rate. Fast rates are achievable in solid-state quantum dot systems, where one can attain a $g/2\pi$ = 20 GHz  and $\kappa / 2\pi$ = 6 GHz \cite{Arakawa2012}, and in which the dipole decay rate can be as low as $\gamma/2\pi$ = 0.16 GHz.  These numbers correspond to a cooperativity of C = 416, which would provide a maximum fidelity of 99.5\%.  The pulse duration of the input photon in the strong coupling regime is limited by the coupling strength to $\tau =1/(2 \pi g)$ = 8 ps, thus providing the possibility for both high fidelity and gigahertz bandwidths.

\section{Serialized qudit-based error correcting teleportation}
We now review QECCs using qudits~\cite{Gottesman1997,Nielsen2010,Gottesman1999}. Errors consist of two types: erasures, which either move qudits out of their $D$-level Hilbert space or reset the system to a particular specified state, and gate errors, which consist of unitary operations acting within the qudits' Hilbert space. These unitary operations may in general act on multiple qudits; however, it is assumed that the action on one qudit is uncorrelated from the action on any other qudit.  This assumption ensures that multiple qudit errors may be characterized using tensor products of single qudit unitary operations. These single qudit unitary operations may be constructed from members of the generalized Pauli group $\mathcal{P}$ \cite{Gottesman1999}, defined as
\begin{align}
\mathcal{P}\equiv\{\sigma_{a,b}=\omega^{c} X^a Z^b;(a,b)\in\mathbb{N}_D^2,c\in\frac{1}{2}\mathbb{N}_{2D}\},
\end{align}
where
\begin{subequations}
\begin{align}
\omega&=e^{2\pi i/D},\\
Z&=\sum\limits_{j=0}^{D-1}\omega^j\ket{j}\bra{j},\\
X&=\sum\limits_{j=0}^{D-1}\ket{(j+1)\, \text{mod}\, D}\bra{j}.
\end{align}
\end{subequations}
Multi-qudit Pauli operations on $n$ qudits are composed of $\mathcal{P}^{\otimes n}$, with the number of nontrivial single qudit operations defined as the weight of the operator. Another single qudit operator outside of the Pauli group but required for this protocol is the higher dimensional analog of the Hadamard gate, the $R$ gate, defined as
\begin{align}
R=\sum_{j,l=0}^{D-1}\omega^{jl}\ket{j}\bra{l}.
\end{align}
An operator that is sufficient to entangle two qudits is the CPHASE gate, which produces a differential phase shift contingent on the state of both qudits:
\begin{align}
\text{CPHASE}=\sum_{j,l=0}^{D-1}\omega^{jl}\ket{j}\bra{j}\otimes\ket{l}\bra{l}.
\end{align}
The CPHASE gate may be raised to any power $q\in\mathbb{N}_D$ to produce a related two qudit gate as well.

An $[[n,k,d]]_D$ stabilizer code is a QECC for $D$-level qudits \cite{Gottesman1997} (when $D$ is omitted, it is to be assumed that $D=2$ and the code is referring to qubits). Here, $n$ is the number of physical qudits utilized for the code, and $k$ is the number of logical qudits they represent. The collection of $D^k$ states making up the QECC are called codewords. The last parameter, $d$, is the lowest weight of any Pauli operator that projects one codeword onto a different codeword. There are $k$ logical phase and bit flip operators, $\{\widebar{Z}_j\}$ and $\{\widebar{X}_j\}$ respectively, associated with the $k$ logical qudits. These operators, which are not unique due to representative freedom for the $k$ logical qudits using the codewords, obey the same commutation relations as single qudit Pauli operations. Finally, there are $(n-k)$ measurements, belonging to the Abelian stabilizer group $\mathcal{S}$, used to diagnose errors and which commute with each other,  $\{\widebar{Z}_j\}$, and $\{\widebar{X}_j\}$ \cite{Gottesman1997}.

For $q$ located (erasure) and $l$ unlocated (gate) errors to be correctable using a stabilizer code, any two such errors must be distinguishable from one another when they act on any two codewords:
\begin{align}
\bra{\widebar{w}_j} E_{\beta,(q,l)}^\dagger E_{\alpha,(q,l)} \ket{\widebar{w}_m}=C_\mathcal{\alpha\beta}\delta_{jm},
\end{align}
where $\ket{\widebar{w}_j}$ are codewords of the QECC and the $E$ operators possess the prescribed number of errors. This is guaranteed whenever the composite operator $E_{\beta,(q,l)}^\dagger E_{\alpha,(q,l)}$ has total weight less than $d$. Because any two such error operations must have support on the same subset of $q$ physical qudits, the combined weight of the operator is $2l+q$. For a correctable error, the permissible ranges for located and unlocated errors is then
\begin{subequations}
\begin{align}
0\leq l&\leq(d-1)/2,\\
0\leq q&\leq (d-1)-2l.
\end{align}
\end{subequations}
Setting $q$ or $l$ to zero reproduces the familiar results for having all located or unlocated errors \cite{Gottesman1997}.

\begin{figure*}[t!]
\includegraphics[width=\textwidth]{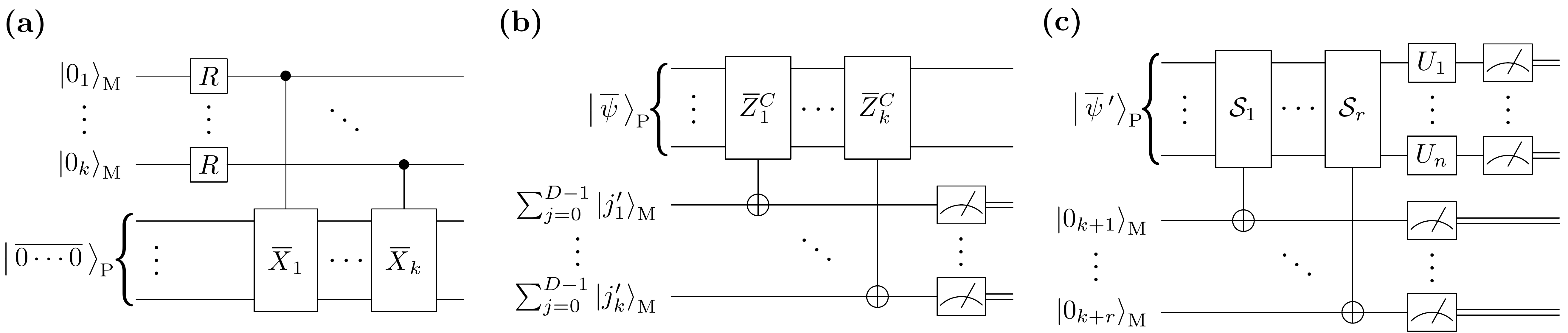}
\caption{\label{fig:jointgen}General quantum circuit to be used at one node of the serial teleportation scheme, broken down into sections. Primes ($^\prime$) denote a state that is entangled with other qudits that are not pictured. (a) entangles the outgoing $\ket{\,\protect\widebar{0\cdots0}\,}$ logical state with $k$ matter qudits and produces an equal weight superposition of every logical codeword entangled with the matching set of logical matter qudit states. (b) entangles the incoming logical state $\ket{\,\overline{\psi}\,}$ with the already entangled $k$ matter qudits. Measurements of the matter qudits are performed, with the results being fed forward, and codewords contained in  $\ket{\,\overline{\psi}\,}$ are left entangled with the codewords from the outgoing states. (c) demonstrates the measurement of the stabilizer generators on the incoming photons. The first $r$ ``expensive'' stabilizers are measured using QND measurements on ancilla matter qudits. The final $(n-k)-r$ ``cheap'' stabilizers are measured by individually measuring the photons, which is permissible because the ``cheap'' stabilizers commute at the individual qudit level. This step simultaneously measures the stabilizers and checks for erasure errors, with results fed forward for Pauli frame correction at the last node.}
\end{figure*}

In order to perform measurements in $\mathcal{S}$, we focus on using quantum non-demolition (QND) measurements. $Z_j^q$ QND measurements are performed using an $R$ gate applied to an ancilla qudit, followed by a $\text{CPHASE}^q$ gate with the $j$th physical qudit and the ancilla, proceeded by an $R^{-1}$ gate on the ancilla. Likewise, $X_j^q$ measurements may be performed by first performing an $R$ gate on both the physical and ancilla qudit, performing a $\text{CPHASE}^q$ gate between them, and rotating both the physical and ancilla qudits back to the original basis using two $R^{-1}$ gates. Once all operations from a stabilizer have been performed using the same ancilla qudit, this ancilla is measured in the $Z$ basis to complete the stabilizer measurement. If, however, the state is not needed anymore, a stabilizer measurement may be performed destructively using measurements directly on the $n$ physical qudits. Furthermore, if multiple stabilizers are strictly composed of single qudit operators that mutually commute, they may be measured simultaneously in this manner.

Creation of an outgoing photon in state $\ket{0}$ is replaced with generation of the QECC codeword representing all zeros for the $k$ logical qudits, $\ket{\,\widebar{0\cdots0}\,}$. Efficiently generating codewords of QECCs has been a topic of interest since the inception of QECCs \cite{Steane1997,Steane2002,Reichardt2004,Paetznick2011}. The schemes that are permissible for a serial approach become even more restrictive, owing to the requirement of at most one interaction between any photon and matter qudit and the corresponding prohibition of multi-photon gates.

One can simply measure all stabilizers with $(n-k)$ matter qudits to produce a codeword. However, we would prefer to use fewer, For CSS codes, Steane's Latin rectangle method \cite{Steane2002} and related techniques \cite{Paetznick2011} are used to efficiently produce the outgoing state with little overhead in non-serialized settings. These approaches would rely upon using photon-photon gates, and thus are not generally useful to this serial teleportation scheme. However, Steane's procedure of employing the generator matrix of one of the QECC's dual classical codes as a gate map is still viable, except that each line of the matrix must be used as a map for gates with an ancilla dot. In this manner, a $\ket{\,\widebar{0\cdots0}\,}$ codeword may be created using $<(n-k)$ stabilizers in general and $(n-k)/2$ stabilizers for the particular codes considered here.

%

Entanglement generation is adjusted to accomodate the $k$ logical qudits of the QECC by employing $k$ matter qudits. Using each matter qudit as a control qudit and following the same guidelines used for imprinting stabilizer measurements on ancillas, each operator $\widebar{X}_j$ is performed as a controlled operation. This is demonstrated in panel (a) of Fig.~\ref{fig:jointgen}. The incoming logical state may then be entangled with the $k$ matter qudits, this time using the $\{\widebar{Z}_j\}$ operators  to determine which gates to apply. Because the photons are used as controls rather than targets in this case, these phase controlled flip operations undergo the substitutions
\begin{subequations}
\begin{align}
Z_j^q&\rightarrow\text{R}_{\text{M}}^{-1}\,\text{CPHASE}_j^{D-q}\,\text{R}_{\text{M}},\\
X_j^q&\rightarrow\text{CPHASE}_j^{D-q}. 
\end{align}
\label{eq:enc}
\end{subequations}
The $j$ subscript now refers to the photonic control qudit, and $R$ gates are now performed on the matter qudits rather than the photons. These new operators controlling the matter qudits are called $\{\widebar{Z}_j^C\}$. The entangling operations are followed by measurements in the $Z$ basis of every matter qudit with results fed forward for Pauli frame correction \cite{Knill2005a,Aliferis2006}. The circuit required to carry out this procedure is pictured in panel (b) of Fig.~\ref{fig:jointgen}.

Rather than strictly completing a bell measurement, this is the point at which stabilizer generators for the QECC must be measured. Constructing subsets of the stabilizers which consist of strictly commuting single qudit Pauli operators, selecting the subset that maximizes the amount of nontrivial gates therein, and selecting these $(n-k)-r$ stabilizers in the subset to be measured destructively from the individual photons minimizes the number of QND stabilizer measurements required. From this standpoint, it is beneficial to select QECCs that have large subsets of stabilizers that commute at the individual qudit levels, such as CSS codes. The first $r$ stabilizers are then measured using QND measurements, the final $(n-k)-r$ stabilizer measurements are inferred from the destructive photon measurements along with the locations of any missing qudits, and the results are fed forward for Pauli frame correction. This stabilizer measurement step is pictured in panel (c) of Fig.~\ref{fig:jointgen}.

Finally, amassing the results of measurements made to create $\ket{\,\widebar{0\cdots0}\,}$, measurements of the $k$ matter qudits, the QND measurements used to measure the first $r$ stabilizers, and the destructive measurements of all $n$ photons at each repeater node prescribes one final error correction step to be made after the last node. Due to the use of state teleportation, gate and erasure errors on individual photons manifest themselves as errors in the logical codeword basis of the outgoing photons and can be understood as a change of the Pauli frame. As long as no \emph{individual} node registers more than the permitted amount of located and unlocated errors, the final output state is guaranteed to be correctable.

\section{Numerical Analysis of Performance}
\label{sec:num}
We now consider optimization of this protocol for a realistic set of errors. We will consider approaches to maximize the entangled bit rate per photon generated by optimizing the number of repeater nodes per attenuation length denoted by the repeater density $\eta$. Successful entangled bit generation using this serialized teleportation protocol hinges on each node registering a correctable subset of errors. The optimal repeater spacing $\ell$ maximizing this rate is determined by balancing photon losses during the transmission between nodes with the erasure and gate errors that may occur in components used for the serial repeater. Using this optimal repeater spacing, the performance of the serial repeater protocol for various QECC assuming different error rates may be calculated and compared to sending single photons down a lossy transmission line without a repeat, henceforth called ``bare transmission.'' First, an error model must be defined to quantify this performance. The following assumptions are made about the operation of components used within the protocol:
\begin{enumerate}
\item All single qudit operations (other than photon propagation and photon measurement) are assumed to be perfect.
\item Whenever missing qudits were supposed to be acted upon by a logical gate, that gate instead behaves as an identity operation, leaving the photon missing as before and the state of any control or target ancilla qudit unchanged.
\end{enumerate}
The errors that may occur within the different components utilized for the protocol may be split into locatable erasures and unlocated gate errors. Locatable erasure errors include:
\begin{enumerate}
\setcounter{enumi}{2}
\item During photon generation, there is a probability $p_C$ that no photon is created. However, if a photon is produced, it is precisely the intended state without logical error.
\item At two qudit logical gates, the photon may interact with the matter qudit as intended, but go missing with probability $p_G$ \emph{after} completing a successful logical operation.
\item During transmission between nodes, we assume the photon is free of logical errors. However, the photon encounters a constant probability to be lost from the fiber. This gives a probability $u=\exp(-\alpha\ell)$ of traveling a distance $\ell$, where $\alpha$ is the characteristic rate of photon loss with distance.
\item The photon may hit a detector without registering a measurement, producing an erroneous null result with probability $p_M$.
\end{enumerate}
Components that may introduce unlocated errors include:
\begin{enumerate}
\setcounter{enumi}{6}
\item Two qudit gates produce a Pauli group error on the photon with probabilities $p_X$, $p_Y$, and $p_Z$ for bit flip errors, simultaneous bit and phase flip errors, and phase flip errors, respectively.
\item Photons may hit a detector while registering an erroneous readout for the measurement with probability $p_F$.
\end{enumerate}
Finally, detector dark counts may manifest as unlocated errors, reproduce the errorless syndrome, or prevent erasure errors from registering:
\begin{enumerate}
\setcounter{enumi}{8}
\item Dark counts are assumed to occur independent of whether a photon was present at the detection stage. $D-1$ out of $D$ times, this dark count will register as a logical error. One out of every $D$ times, it produces a syndrome that will yield the proper recovery operation.
\end{enumerate}

\begin{figure*}[t!]
\includegraphics{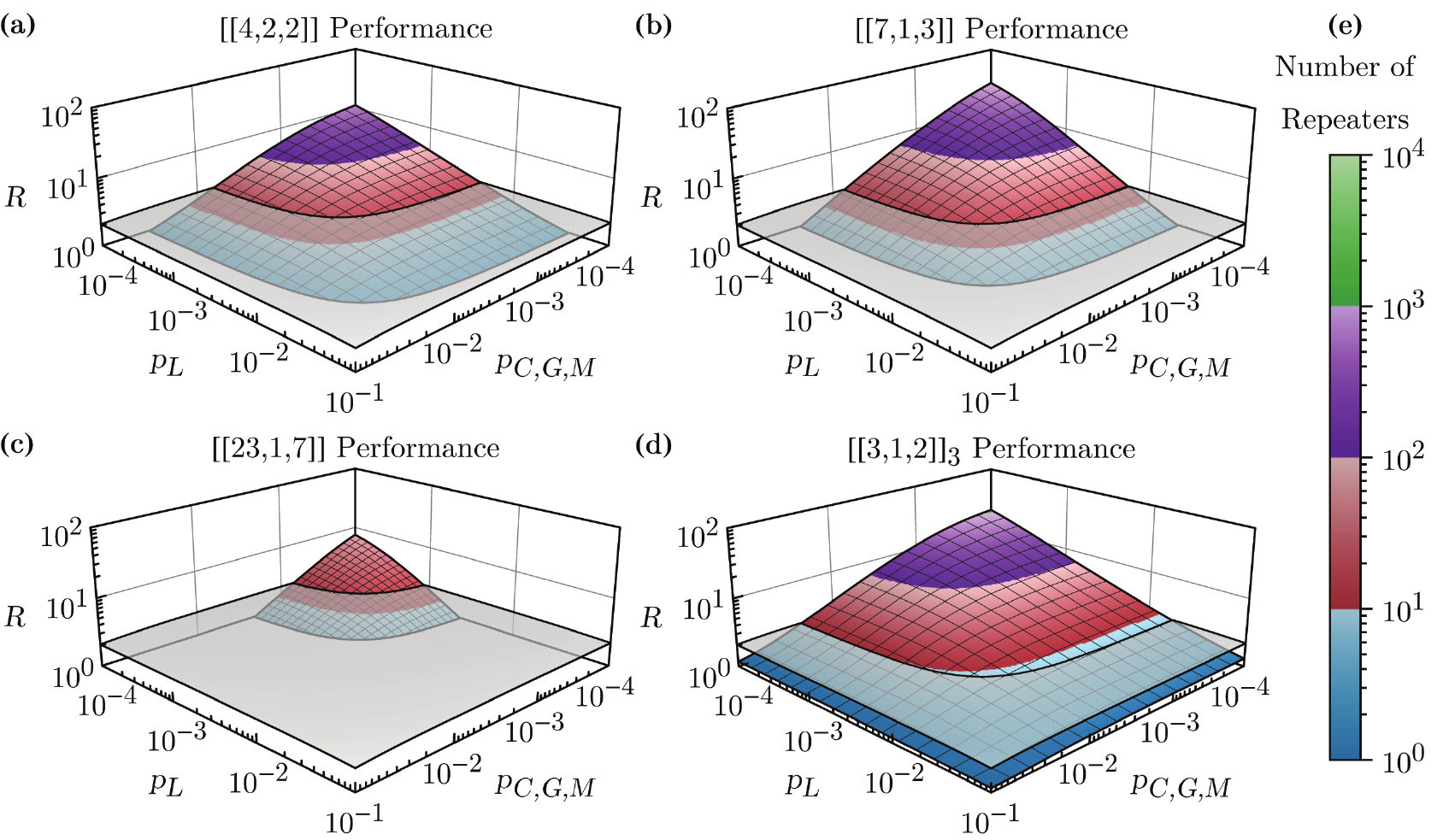}
\caption{\label{fig:plots}Plots of the performance of the serialized repeater scheme compared to bare transmission for various CSS codes, namely the (a) $[[4,2,2]]$, (b) $[[7,1,3]]$, (c) $[[23,1,7]]$, and (d) $[[3,1,2]]_3$ codes. Coloring of the surfaces is done to highlight the number of repeater nodes required to reach that indicated $R$ value, with the color scale given by (e). $P_{\text{tot}}$ values have been fixed at $0.1$, $P_{DC}$ has been set to zero, $P_F$ has been set to $2P_L/3$, and $P_C$, $P_G$, and $P_M$ have been set equal.  The max achievable distance for bare transmission for the given $P_{\text{tot}}$ is plotted in gray on every plot.}
\end{figure*}

From these assumptions, a performance function for the operation of an $[[n,k,d]]_D$ based node may be constructed, giving the probability that the output state will produce a correctable outcome after traveling a distance $\ell$ and being repeated at a single node. In this case, it proves beneficial to use the probability $u$ rather than the distance $\ell$. This single node performance function requires the enumeration of all possible ways in which the number of located and unlocated errors is below the threshold for correctability. Each individual qudit may encounter a different number of logical gates dependent on the circuit used for the given QECC, so in general this requires analyzing each photonic qudit's probability of success separately and combining the results such that the condition for correctability is met. Defining the probability of receiving the $j$th qudit without error, the probability of producing a correctable unlocated error, and the probability of the qudit having gone missing as $P_{A,j}(u)$, $P_{B,j}(u)$, and $P_{C,j}(u)$ respectively (see Appendix~\ref{app:A}), a single node performance function may be defined as
\begin{widetext}
\begin{align}
S(u)=\sum_{l=0}^{\lfloor(d-1)/2\rfloor}\quad \sum_{q=0}^{(d-1)-2l}
\sum_{\substack{\text{Permutations of }q\\\text{located and }l\\\text{unlocated errors}}}
\quad\prod_{j=1}^{n} P_{\xi,j}(u),
\label{eq:performance}
\end{align}
\end{widetext}
where $\xi$ takes on the indices $A$, $B$, and $C$ depending on which permutation of errors is examined. This function defines the probability the quantum repeater node will produce a final state from which the correct state is recoverable with a known set of operations.

If there is to be any benefit for using multiple repeaters, there must be some length scale $\ell_>$ defined by $u=\exp(-\alpha\ell_>)$ for which $S(u)$ is greater than the bare transmission case. This will always be possible as long as
\begin{align}
S(u)>u
\end{align}
for some $u\in(0,1)$ provided the error rates lie below the threshold for quantum error correction. In this manner, eliminating for $u$ using the set of equations
\begin{subequations}
\begin{align}
S(u)&=u,\\
\frac{dS}{du}&=1
\end{align}
\end{subequations}
provides an equation defining a ``threshold relationship'' between all of the error probabilities. Given a set of probabilities that solve this threshold relationship, one may guarantee that the serial protocol is better than bare transmission for some value of $u$ by decreasing any of the error probabilities by any amount.

For transmission over long distances, the use of multiple repeaters will provide a benefit to achievable success rates as long as error probabilities lie below the threshold for error correction. It is then pertinent to optimize the distance $\ell$ between adjacent nodes to maximize the chance of successfully teleporting a state over some longer distance $L$. Working with the dimensionless parameter $R=\alpha L$ (representing the ratio of the length $L$ to the attenuation length of a photon in the optical fiber), this amounts to maximizing the function
\begin{align}
P_{\text{tot}}(u)=\left[S(u)\right]^{-\frac{R}{\log u}},
\label{eq:ptot}
\end{align}
where the exponent represents the (approximate) number of nodes covering the distance $L$. The value of $u$ that maximizes $P_{\text{tot}}$ is independent of $R$ (and vice versa) and may be found numerically by solving the transcendental equation
\begin{align}
S(u)\log\left[S(u)\right]=u\frac{dS}{du}\log u
\label{eq:trans}
\end{align}
for $u$. An equivalent form of Eq.~(\ref{eq:trans}), useful for short distances when $u$ is expected to be very close to $1$,  may be written as an equation for $\epsilon=(1-u)$:
\begin{widetext}
\begin{align}
\ln\left[S(1)\right]=\sum_{j=2}^{\infty}\frac{\epsilon^j}{j!} \Bigg[(-1)^j(j-1)\frac{d^j}{du^j}\left[\ln S(u)\right]
+\sum_{m=0}^{j-2}\binom{j}{m}(j-m-2)!(-1)^m\frac{d^{m+1}}{du^{m+1}}\left[\ln S(u)\right]\Bigg]\Bigg|_{u\rightarrow 1}.
\label{eq:expansion}
\end{align}
\end{widetext}
In general, higher order terms may be neglected when the error rates are very low and provided higher order derivatives of $S(u)$ are of $\mathcal{O}(1)$ near $u=1$.

Solving Eq.~(\ref{eq:ptot}) for $R$ and evaluating it at the value $u$ solving Eq.~(\ref{eq:trans}) provides an output representing the furthest possible (dimensionless) distance at which one could still expect to receive the correct quantum state at the last node with probability $P_{\text{tot}}$. Assuming that approximations to Eq.~(\ref{eq:expansion}) may be made to lowest order, $R$ may be estimated as
\begin{align}
R\approx\frac{\sqrt{2}\, S(1) \abs{\ln P_{\text{tot}}}}{\left[\abs{\ln S(1)}\left[S^\prime(1)^2-S(1)\left(S^{\prime\prime}(1)+S^\prime(1)\right)\right]\right]^{\frac{1}{2}}}
\label{eq:approxR}
\end{align}
for a repeater density $\eta=1/\alpha\ell$ of
\begin{align}
\eta\approx\frac{1}{S(1)}\sqrt{\frac{S^\prime(1)^2-S(1)\left(S^{\prime\prime}(1)+S^\prime(1)\right)}{2\abs{\ln S(1)}}}.
\label{eq:lapprox}
\end{align}
Eqs.~(\ref{eq:approxR}) and (\ref{eq:lapprox}) are useful for finding limiting behavior once the full form of $S(x)$ is known for a given QECC. To lowest order, $R$ depends on error probabilities through a power law relationship, but this behavior breaks down when the error rate being varied is of the same order as other fixed error probabilities.

\begin{figure*}[t!]
\includegraphics{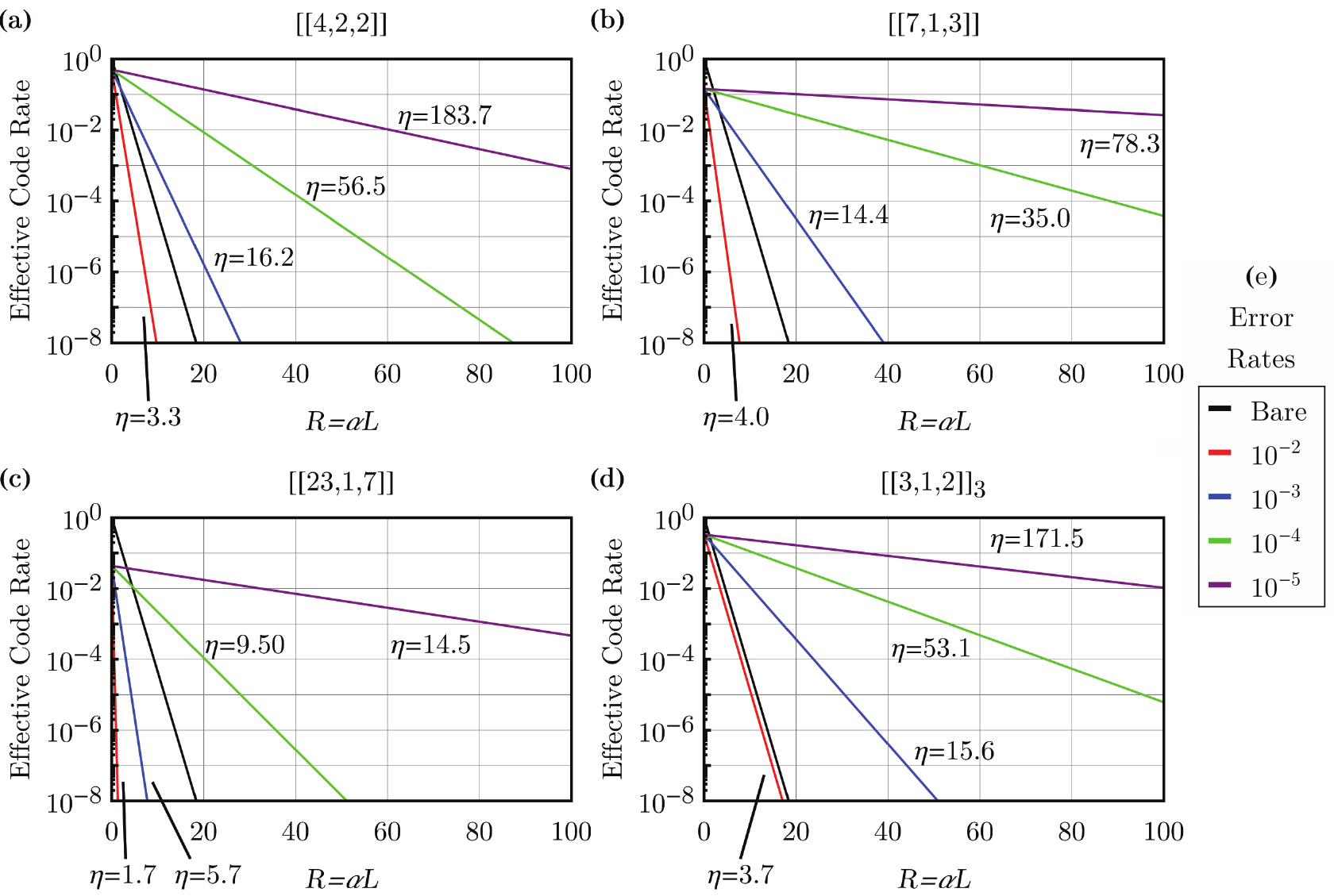}
\caption{Plot of effective code rate versus unitless distance $R=\alpha L$ for the (a) $[[4,2,2]]$, (b) $[[7,1,3]]$, (c) $[[23,1,7]]$, and (d) $[[3,1,2]]_3$ codes. Effective code rate is defined as $P_{\text{tot}}\times k/n$. The colored lines correspond to different error rates of individual gates as pictured in (e), with the black line corresponding to bare transmission. The error rates listed correspond to the values of every error probability $3P_F/2=P_L=P_C=P_G=P_M$. Dark counts are neglected here. Multiplying the probability values here by the rates in Table~\ref{tab:perf} and $n$ yields true entangled bit/second rates at the specified distance and error rate.\label{fig:rates}}
\end{figure*}

The analysis here is restricted to CSS codes due to the aforementioned property of having cheap stabilizer measurement costs and simple codeword generation schemes, as documented in Table~\ref{tab:perf}. Four codes are analyzed in both Fig.~\ref{fig:plots} and Fig.~\ref{fig:rates}. The $[[4,2,2]]$ (panel (a)) and $[[7,1,3]]$ (panel (b)) codes are chosen because they are the simplest CSS qubit codes that correct one and two errors, respectively. The $[[23,1,7]]$ (panel (c)) Golay code is chosen because it has been shown to be an excellent performing QECC in other work \cite{Steane2003}. Finally, the $[[3,1,2]]_3$ (panel (d)) qutrit code has been chosen both because it is the simplest QECC that can correct one error using three qudits, and because it has an extremely simple generating circuit and set of stabilizers.

\begin{table}[b]
\begin{tabular}{c|cccc}
\textbf{Codes} & \makecell{\textbf{Matter} \\ \textbf{qudits} } & \makecell{\textbf{Elements used} \\ \textbf{for encoding}} & \makecell{\textbf{Est. ebit} \\ \textbf{rate (GHz)}} & \makecell{\textbf{Eff. att.} \\ \textbf{length  ($1/\alpha$)}} \\
\hline
$[[4,2,2]]$ & 3 & 1 & 2.8 & 1.58 \\
$[[7,1,3]]$ & 4 & 3 & 1.8 & 2.37 \\
$[[23,1,7]]$ & 12 & 11 & 0.6 & 0.51 \\
$[[3,1,2]]_3$ & 2 & 1 & 3.5 & 2.94
\end{tabular}
\caption{Relevant parameters regarding code performance. Listed are the total number of matter qudits required for state teleportation and stabilizer measurements, the number of single photon-on-demand emitters and matter qudits required for outgoing $\ket{\,\protect\widebar{0\cdots0}\,}$ preparation, and the maximum estimated entangled bit rates rates achievable if $P_{\text{tot}}=1$ by assuming 100 ps per photon gate for quantum dot-based approaches as discussed in the main text. These base rates are calculated assuming QND measurements may be performed on the matter qudits using, e.g., CPHASE gates with two photons and multiplexed high-efficiency photon detection. Effective attenuation lengths are calculated under the same assumptions used in Fig.~\ref{fig:rates} and with a joint error rate of $10^{-3}$.\label{tab:perf}} 
\end{table}

Fig.~\ref{fig:plots} examines the maximum distances achievable given a fixed entangled bit rate over a wide range of error probabilities for the selected codes. $P_{\text{tot}}$ was selected to have a value of $0.1$, $P_{DC}$ was set to zero, and $P_C$, $P_G$, and $P_M$ were set equal to one another. The probabilities for logical errors in gates were set equal and to a sum total of $P_L$ (i.e. $p_X=p_Y=p_Z=P_L/3$). Finally, $p_F$ was set to a value of $2P_L/3$. Three dimensional plots were produced, plotting the obtained $R$ value versus both the total logical error probability $P_L$ and the combined erasure errors $P_{C,G,M}$. Also plotted is a surface representing the bare transmission distance at an identical $P_{\text{tot}}$ value. The surfaces for the serial protocol are colored according to how many repeaters are required to reach the plotted $R$ value. The number of repeaters utilized may be decreased, but this will naturally also decrease the $R$ value as the function will no longer be maximized.

The plots in Fig.~\ref{fig:plots} demonstrate improvement in the achievable distance at a fixed rate compared to bare transmission for low error probabilities. The larger the codes, the more it becomes beneficial to space out the nodes, both because the larger distance codes can handle more photon loss in transmission and because the more nodes are used, the more chance for uncorrectable errors to occur during stabilizer measurements. The $[[7,1,3]]$ code demonstrates the best performance of the 4 codes at the lowest error probabilities plotted, with the $[[3,1,2]]_3$ code possessing the least stringent error threshold.

The scaling of $P_{\text{tot}}\times k/n$ with distance $R$ is plotted in Fig.~\ref{fig:rates} at a variety of joint error rates. These error rates correspond to the values of both the logical and erasure error probabilities, with $3P_F/2=P_L=P_C=P_G=P_M$ representing the listed joint rates. Dark count rates are again neglected. True entangled bit/second rates for a QECC at a given distance for the different error probabilities may then be found by taking values from Table~\ref{tab:perf}, multiplying by $P_{tot}\times k/n$ values taken from Fig.~\ref{fig:rates}, and finally multiplying by $n$. The rate lines are evaluated for their optimum repeater spacings $\eta$ found using the $u$ values solving Eq.~(\ref{eq:trans}), with this optimal value listed adjacent to its respective curve. In both Fig.~\ref{fig:plots} and Fig.~\ref{fig:rates}, it is evident that the larger the code, the greater the dependence of both $R$ and $P_{\text{tot}}$ (assuming the other parameter is fixed) on the error probabilities. Limits of performance for a QECC are determined by the number of entangling gates between the incoming photonic qudits and the matter qudits used to establish entanglement and measure the ``expensive'' stabilizers, as any errors in these gates that do not commute with the operations will produce erroneous syndromes (see Appendix~\ref{app:A} for further discussion).

\section{Summary}
We have shown that quantum repeaters can mediate communication within a quantum network using a completely serialized approach, where photons encoded in a quantum error correcting code interact once per photon with each matter qudit in a repeater node before being detected with a near unit-efficiency photodetector. This scheme is particularly well suited to achieving high bit rates and does not require long-term quantum memory. Looking at available physical systems, high bandwidth single photon generation and photon-matter qudit gates remain challenging but rudimentary demonstrations indicate that near giga-entangled bit/second rates may be achieved for near-future device performance. We note that CSS codes seem optimal for serialized approaches because of their simple codeword preparation and the permissibility of joint stabilizer generator measurements. This does not preclude non-CSS codes from also performing well. However, in a comparison between a CSS qubit code and a general non-CSS stabilizer code with similar $n$ and $k$ parameters, the CSS code will have lower overhead afforded by the two aforementioned properties.  Finally, we emphasize the crucial challenge for this approach to building a quantum repeater: one needs extremely low loss, end-to-end single photon generation, transmission, CPHASE gates, and detection. While individual demonstrations of such performance now exist, integrating these pieces into a comprehensive package remains a formidable task.

\begin{acknowledgements}
We thank Sae Woo Nam, Glenn Solomon, Alan Migdall, Rich Mirin, and Liang Jiang for helpful conversations.
This work was supported by DARPA Quiness and by the NSF-funded Physics Frontier Center at the JQI.
\end{acknowledgements}

\bibliography{Graduate_Work-Serial_Encoded_QECC-3}

\appendix

\section{Performance Function}
{\setcounter{equation}{0}
\renewcommand{\theequation}{A\arabic{equation}}
\setcounter{figure}{0}
\renewcommand{\thefigure}{A\arabic{figure}}
\label{app:A}

The performance function defined in Eq.~\ref{eq:performance} depends on the probabilities of the physical qudits to have three different outcomes: to be measured without incurred error, to be measured with a registered gate error, and to not register a photon (null measurement) with the photon detector. The probabilities for the $j$th qudit to produce one of these outcomes are denoted by $P_{A,j}$, $P_{B,j}$, and $P_{C,j}$, respectively.

For a qudit to be measured without any incurred errors, it must proceed through every element of the protocol without any gate or erasure errors. There are, however, a few exceptions to this rule for the CSS codes  considered here. In particular, once the stabilizer measurements for detecting bit flips have been completed, bit flips may occur on the incoming qudits without introducing any error to the outgoing logical state. Furthermore, even if a phase error occurs (or a bit flip/missing qubit error after the bit flip-detecting stabilizers), a dark count can produce an output syndrome indicating that the node has functioned without error. That can yield the proper syndrome for recovery $1$ out of every $D$ times. Thus:
\begin{widetext}
\begin{multline}
P_{A,j}(u)=(1-p_C)(1-p_G-p_X-p_Y-p_Z)^{N_{\gamma,j}+N_{\delta,j}-1}u(1-p_G-p_X-p_Y)(1-p_M-p_F)(1-p_{DC})\\+(1-p_C)(1-p_G-p_X-p_Y)^{N_{\gamma,j}+N_{\delta,j}-1}u\frac{p_{DC}}{D},
\label{eq:norm}
\end{multline}
\end{widetext}
where $N_{\gamma,j}$ and $N_{\delta,j}$ are the number of gates used in the outgoing codeword preparation and entanglement steps and the incoming entanglement and stabilization steps, respectively.

For a qudit to register a correctable gate error, at least one gate error needs to occur somewhere during the operation of the repeater. In general, errors \emph{cannot} occur and be correctable if they happen between two operations with which the error does not commute. For example, if a particular qudit is needed for two different stabilizer measurements that help detect bit flip errors, a bit flip error occurring on this qubit in between the two measurements will produce a syndrome that yields an incorrect recovery operation. This is due to a limited ability to correct errors occurring in certain parts of the circuit. Specifically, in ideal quantum error correction, errors are assumed to be incurred during operation of some quantum circuit, and \emph{perfect} error detection is able to detect and correct them afterwards. Accounting for errors happening in the stabilizer measurements themselves creates scenarios where a particular (normally correctable) syndrome may be produced either by errors during the operation of the quantum circuit or in an error during the stabilizer measurements, each requiring different recovery operations. In this case, the simple strategy adopted here is to assume that the most likely scenario occurred, i.e. that the error happened during operation of the quantum circuit.

Using this assumption, we cannot correct bit flip errors occurring during the entangling gates between the incoming photons and the dots used to mediate state teleportation as well as the dots used for stabilization. Furthermore, during these gates in which some gate errors are detrimental to proper operation, it is still permissible to incur any error that commutes with all operations either before or after the error occurred.

In addition, dark counts that occur when a qudit has otherwise undergone errorless operation, or dark counts that occur when a qudit is in fact lost, may register as gate errors. In the case when a qudit has gone missing (as long as it has not gone missing \emph{between} stabilizer operations that would detect the change),  dark counts  produce a syndrome that  appears to be an unlocated gate error. For the case of otherwise errorless operation with a dark count, the additional error will appear to be an unlocated gate error whenever the result is precisely any dark count result different than those accounted for in Eq.~(\ref{eq:norm}). Incorporating these additional errors, and subtracting off the dark counts producing errorless operation, yields 
\begin{widetext}
\begin{multline}
P_{B,j}(u)=(1-p_C)(1-p_G)^{N_{\gamma,j}+1}u(1-p_G-p_X-p_Y)^{N_{\delta,j}-1}(1-p_M)\\-(1-p_C)(1-p_G-p_X-p_Y-p_Z)^{N_{\gamma,j}+N_{\delta,j}-1}u(1-p_G-p_Y-p_Z)(1-p_M-p_F)(1-p_{DC})\\-(1-p_C)(1-p_G-p_X-p_Y)^{N_{\gamma,j}+N_{\delta,j}-1}u\frac{p_{DC}}{D}+p_{DC}-(1-p_C)(1-p_G)^{N_{\gamma,j}}u\, p_{DC}\\+(1-p_C)(1-p_G)^{N_{\gamma,j}}u(1-p_G-p_X-p_Y)^{N_{\delta,j}-1}p_G p_{DC}\\+(1-p_C)(1-p_G)^{N_{\gamma,j}}u(1-p_G-p_X-p_Y)^{N_{\delta,j}-1}(1-p_G)p_M p_{DC}.
\end{multline}
\end{widetext}

Finally, this leaves the probability that the qudit has gone missing in such a way that it is detectable and correctable while registering strictly as a null measurement at the detector. Much like the case for gate errors, erasure errors do not commute with the stabilizer measurement stage. This again prohibits correcting for errors in which the photon goes missing in between these stabilizer measurements and the final photon detection stage. However, as the loss is known, gate errors incurred before the photon has been lost are correctable as well. Parsing out the correctable possibilities, one arrives at the expression
\begin{widetext}
\begin{multline}
P_{C,j}(u)=\Big[1-(1-p_C)(1-p_G)^{N_{\gamma,j}}u+(1-p_C)(1-p_G)^{N_{\gamma,j}}u(1-p_G-p_X-p_Y)^{N_{\delta,j}-1}p_G\\+(1-p_C)(1-p_G)^{N_{\gamma,j}}u(1-p_G-p_X-p_Y)^{N_{\delta,j}-1}(1-p_G)p_M\Big](1-p_{DC}).
\end{multline}
\end{widetext}
Inserting these results into the performance function given by Eq.~(\ref{eq:performance}) for the $n$ qudits permits the analysis in the main text to be carried out.
}

\end{document}